\documentclass[twocolumn,superscriptaddress,epsf,psfig]{revtex4-2}
\usepackage{amssymb}
\usepackage{epsfig}
\DeclareMathAlphabet{\pazocal}{OMS}{zplm}{m}{n}            
\usepackage{graphicx}
\usepackage{color}
\usepackage{bm}
\usepackage[normalem]{ulem}     
\usepackage{soul}
\usepackage{amsmath}
\usepackage{braket} 
\usepackage{rotating}
\usepackage{multirow} 
\DeclareMathAlphabet{\pazocal}{OMS}{zplm}{m}{n}            
\usepackage{graphicx}
\usepackage{hyperref}

\begin{document}
\title{Emergent surface multiferroicity}

\author{Sayantika Bhowal*} 
\affiliation{Materials Theory, ETH Z\"{u}rich, Wolfgang-Pauli-Strasse 27, 8093 Z\"{u}rich, Switzerland} 
\affiliation{Department of Physics, Indian Institute of Technology Bombay, Mumbai 400076, India}
\author{Andrea Urru*}
\affiliation{Materials Theory, ETH Z\"{u}rich, Wolfgang-Pauli-Strasse 27, 8093 Z\"{u}rich, Switzerland} 
\affiliation{Department of Physics \& Astronomy, Rutgers University,
Piscataway, New Jersey 08854, USA}
\author{Sophie F. Weber}
\affiliation{Materials Theory, ETH Z\"{u}rich, Wolfgang-Pauli-Strasse 27, 8093 Z\"{u}rich, Switzerland}
\author{Nicola A. Spaldin}
\affiliation{Materials Theory, ETH Z\"{u}rich, Wolfgang-Pauli-Strasse 27, 8093 Z\"{u}rich, Switzerland}

*These authors contributed equally to the manuscript.
\date{\today}

\begin{abstract}
We show that the surface of a centrosymmetric, collinear, compensated antiferromagnet, which hosts bulk ferroically ordered magnetic octupoles, exhibits a linear magnetoelectric effect, a net magnetization, and a net electric dipole moment. Thus, the surface satisfies all the conditions of a multiferroic, in striking contrast to the bulk, which is neither polar nor exhibits any net magnetization or linear magnetoelectric response. Of particular interest is the case of non-relativistic $d$-wave spin split antiferromagnets, in which the bulk magnetic octupoles and consequently the surface multiferroicity exist even without spin-orbit interaction.
We illustrate our findings using first-principles calculations, taking FeF$_2$ as an example material. Our work underscores the bulk-boundary correspondence in these unconventional antiferromagnets.
\end{abstract}

\maketitle

In the last decades, many examples of unusual behaviors at surfaces and interfaces that differ markedly from the properties of the corresponding bulk materials have been identified. Examples include the formation of conducting or even superconducting layers at the interfaces between bulk band insulators \cite{Reyren2007}, conducting spin-polarized edge states at the surfaces of bulk topological insulators \cite{Hasan-Kane2010}, and the emergence of net magnetization at the surfaces of antiferromagnets (AFMs) \cite{Belashchenko2010, Sophie2023} or at their interfaces with paramagnets \cite{Nichols2016}. These properties are enabled both by changes in chemistry and by symmetry lowering, which permits characteristics that are prohibited in the bulk. An understanding of the emergent properties associated with surfaces is of importance both for interpreting experimental measurements of ostensibly bulk phenomena in finite-sized samples, and for engineering devices, which often rely on a thin-film geometry.

In this work, we reveal the coexistence of net magnetization, electric polarization, and the linear magnetoelectric (ME) effect, at the surface of an AFM material whose bulk symmetry prohibits all three properties. This combination of properties is characteristic of ME multiferroics, defined to be materials which contain the two primary ferroic orders, magnetization and electric polarization. Since ME multiferroics are insulators that break both time-reversal ($\cal T$) and space-inversion ($\cal I$) symmetries, they also exhibit the linear ME effect, in which the magnetization, $m_j$, responds linearly to an external electric field, ${\cal E}_i$, $\Delta m_j=\alpha_{ij}{\cal E}_i$, with $\alpha_{ij}$ the linear ME response tensor, and vice versa \cite{SpaldinFiebig2005}. We note that, while ME multiferroics are intensively researched, combining all three properties in a bulk material is challenging \cite{Nicola2000}, and the multiferroic candidates currently explored for potential novel microelectronic devices are in fact {\it anti-}ferromagnets \cite{Spaldin-Ramesh2019, Spaldin2010}. In this context, the surface multiferroicity introduced here, with its net magnetization, might be particularly relevant for applications.

The emergent surface multiferroicity, that we introduce here, occurs in AFMs that contain ferroically aligned local magnetic octupoles, ${\cal O}_{ijk} = \int \mu_i (\vec r) r_j r_k d^3r$ leading to a net magnetic octupole per unit volume, which we call a magneto-octupolization by analogy with the magnetization and polarization. Here  $i,j,k$ represent the Cartesian directions and $\vec \mu (\vec r)$ is the magnetization density. The ferromagneto-octupolar order in such AFMs breaks $\cal T$ symmetry in spite of the absence of a net magnetic dipole. Since $\cal I$ symmetry is preserved, the bulk linear ME effect is prohibited, however a quadratic ME effect, $\Delta m_i =  \beta_{ijk} {\cal E}_j {\cal E}_k $ is symmetry allowed. Here $\beta_{ijk}$ is the quadratic ME response tensor, with its non-zero components determined by the form of the magnetic octupolar order \cite{Urru2022}. It was recently shown that systems with such ferromagneto-octupolar order have a net surface magnetization as a second order response to the intrinsic electric field of the surface \cite{Sophie2023}. Here we find that in addition to the surface magnetization, the bulk magnetic octupoles also lead to surface ME multipoles ${\cal M}_{ij} = \int r_i \mu_j (\vec r) d^3r$ as a linear response to the intrinsic electric field of the surface, and that these give rise to the subsequent surface linear ME response. The magnetic octupole is therefore the source of the surface multiferroicity. 

We demonstrate the emergent surface multiferroic behavior computationally using rutile-structure FeF$_2$ as our prototype material. Our motivation for this choice is three-fold. First, it has the same crystallographic and magnetic structure as MnF$_2$, which was recently shown to have ferroically ordered magnetic octupoles \cite{BhowalSpaldin2024} with an associated quadratic ME effect. Second, it was used as an example to illustrate the existence of surface magnetization on surfaces of bulk AFMs with compensated magnetic dipoles \cite{Sophie2023}, 
 offering an explanation for the large experimentally observed exchange bias \cite{Lapa2020}. 
Finally, since the magnetic octupoles are not a consequence of spin-orbit coupling (SOC), it exhibits a substantial non-relativistic spin splitting (NRSS) of the electronic bands with a $d$-wave symmetry in reciprocal space, a phenomenon referred to as altermagnetism \cite{Libor2020, Naka2019, Kyo-Hoon2019, Hayami2019, Yuan2020, Yuan2021, Smejkal2022PRX, Yuan2022arXiv, YuanNov2022arXiv}.  

We study FeF$_2$ using density functional theory (DFT) as implemented within the Vienna ab initio simulation package (VASP) \cite{Kresse1993, Kresse1996} along with symmetry-based analysis. Our main finding is that, in contrast to bulk FeF$_2$ which has zero magnetization, polarization and linear ME response, the $(110)$ [$(1{\bar1}0)$] surface of FeF$_2$ is multiferroic with a net electric polarization along the $[110]$ ($[1{\bar1}0]$) direction, a net magnetization along the $[00\bar{1}]$ ($[001]$) direction, and an off-diagonal linear ME response $\alpha_{xz}=\alpha_{yz}$ ($\alpha_{xz}=-\alpha_{yz}$). 

\begin{figure}[t]
\includegraphics[width=\columnwidth]{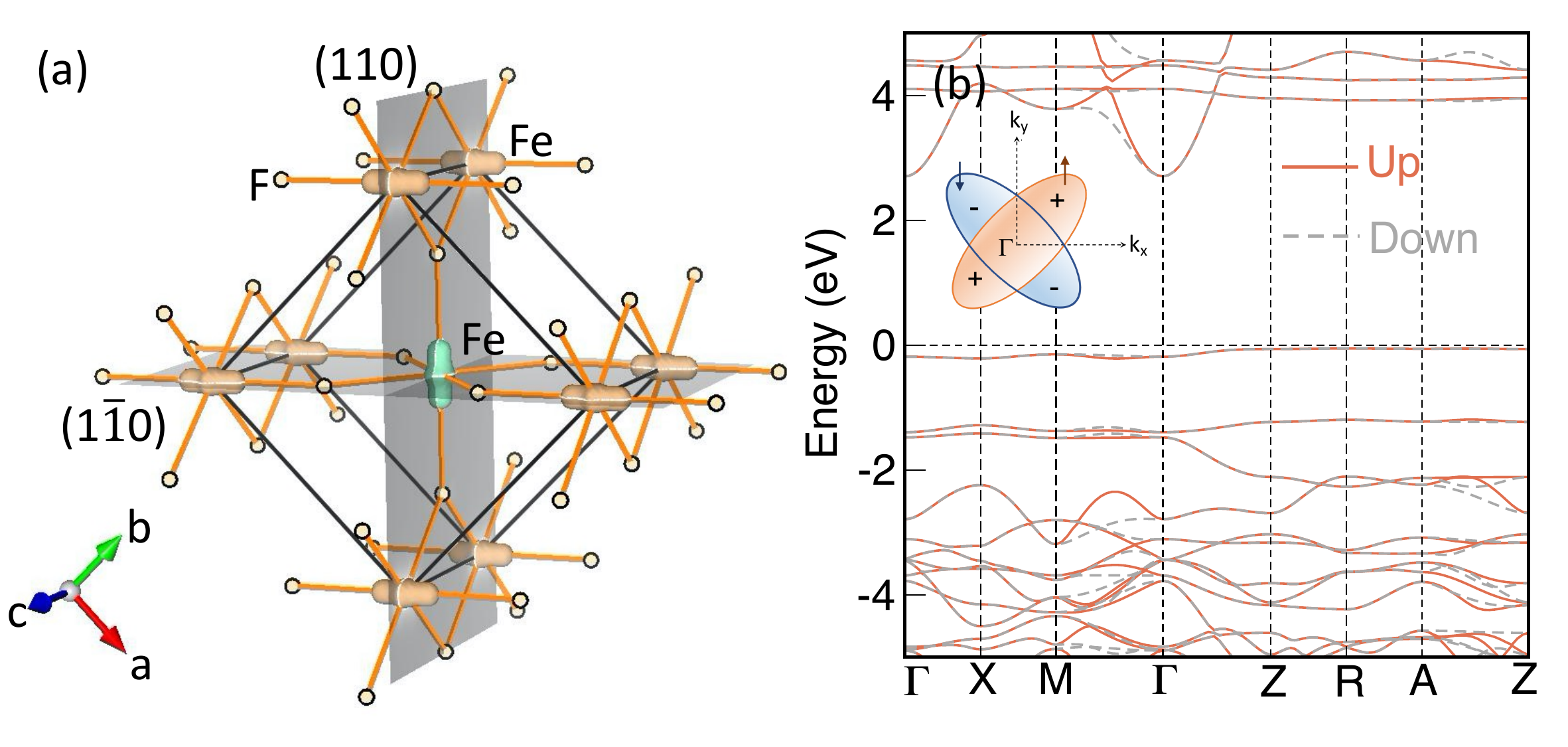}
\caption{(a) Band decomposed magnetization density in bulk FeF$_2$. The (110) and ($1\bar{1}0$) planes of the bulk structure are also indicated. (b) Band structure of bulk FeF$_2$, showing the large energy splitting of the spin-polarized bands with the solid and dashed lines indicating the up and down spin-polarized bands respectively. The inset shows the schematic illustration of the $d$-wave pattern of spin splitting relevant to FeF$_2$.}
\label{fig2}
\end{figure}

{\it Case of FeF$_2$}--
Bulk FeF$_2$ crystallizes in a tetragonal structure with two Fe ions of oppositely oriented magnetic dipole moments with directions along $[001]$ and $[00\bar{1}]$ for the corner and the center Fe ions of the unit cell \cite{Strempfer2004}. 
The nonmagnetic F environments surrounding the two Fe atoms are inequivalent (see Fig. \ref{fig2}a), leading to broken global $\cal T$ symmetry, even though there is no net spin dipole moment in the unit cell. 
As shown in Fig. \ref{fig2}a, the FeF$_6$ octahedral coordination lead to highly anisotropic magnetization density around the Fe ions, indicating the presence of magnetic multipoles beyond the dipolar order due to deviation from the spherical symmetry. 

We therefore perform a magnetic multipole analysis within the DFT+$U$ method for the ground state of bulk FeF$_2$ \cite{note1, Spaldin2013, Bultmark2009}. Our calculations show that the 
first-order asymmetry in the magnetization density, dictated by the ME multipole ${\cal M}_{ij}$, 
vanishes at the Fe ions, consistent with the presence of global as well as local (at the Fe ions) inversion symmetry, and the absence of the linear ME effect in bulk FeF$_2$. We find that components of the next higher-order magnetic multipole, the magnetic octupole, are non-zero and order ferroically, consistent with the symmetry and our previous work on the isostructural MnF$_2$ \cite{BhowalSpaldin2024}. 

The magnetic octupole ${\cal O}_{ijk}$ is a rank-3, $\cal I$ symmetric but $\cal T$ broken magnetic multipole, that describes the second-order asymmetries in the magnetization density. The ${\cal O}_{ijk}$ tensor can be decomposed into different irreducible (IR) components \cite{Urru2022}, among which we find that the totally symmetric component ${\cal O}_{32^-}$ (the subscript indices denote respectively the rank $l$ and component $m$, which can take any value between $-l,...,+l$)
and one of the non-symmetric components, the magnetic toroidal quadrupole moment ${\cal Q}^{(\tau)}_{x^2-y^2}$, have a ferro-type ordering in FeF$_2$. These two components lead to a bulk {\it global} octupole tensor with the non-zero elements ${\cal O}_{xyz}={\cal O}_{yxz}\ne{\cal O}_{zxy}$ in FeF$_2$.
The ferroic ordering of magnetic octupole components gives rise to a large spin splitting of the bands along the $\Gamma \rightarrow M$ direction in the reciprocal space of FeF$_2$ (Fig. \ref{fig2}b), which switches sign under $C_4$ rotation of the momentum direction in a $d$-wave pattern (see inset of Fig. \ref{fig2}b).
In addition to the ferro-type magnetic octupole components at the Fe sites, our calculations show that there are also non-zero magnetic octupole components ${\cal O}_{30}$ and $t_z^{(\tau)}$ which have an antiferro-type arrangement with opposite signs between the Fe ions. These antiferroic magnetic octupole components, in turn, lead to a {\it local} octupole tensor (indicated by the bar lines) with the components $\bar{{\cal O}}_{xxz}=\bar{{\cal O}}_{yyz}, \bar{{\cal O}}_{zxx}=\bar{{\cal O}}_{zyy} ~~\text{and}~~ \bar{{\cal O}}_{zzz}$ in addition to the global components discussed above.

\begin{figure*}[t]
\includegraphics[scale=0.25]{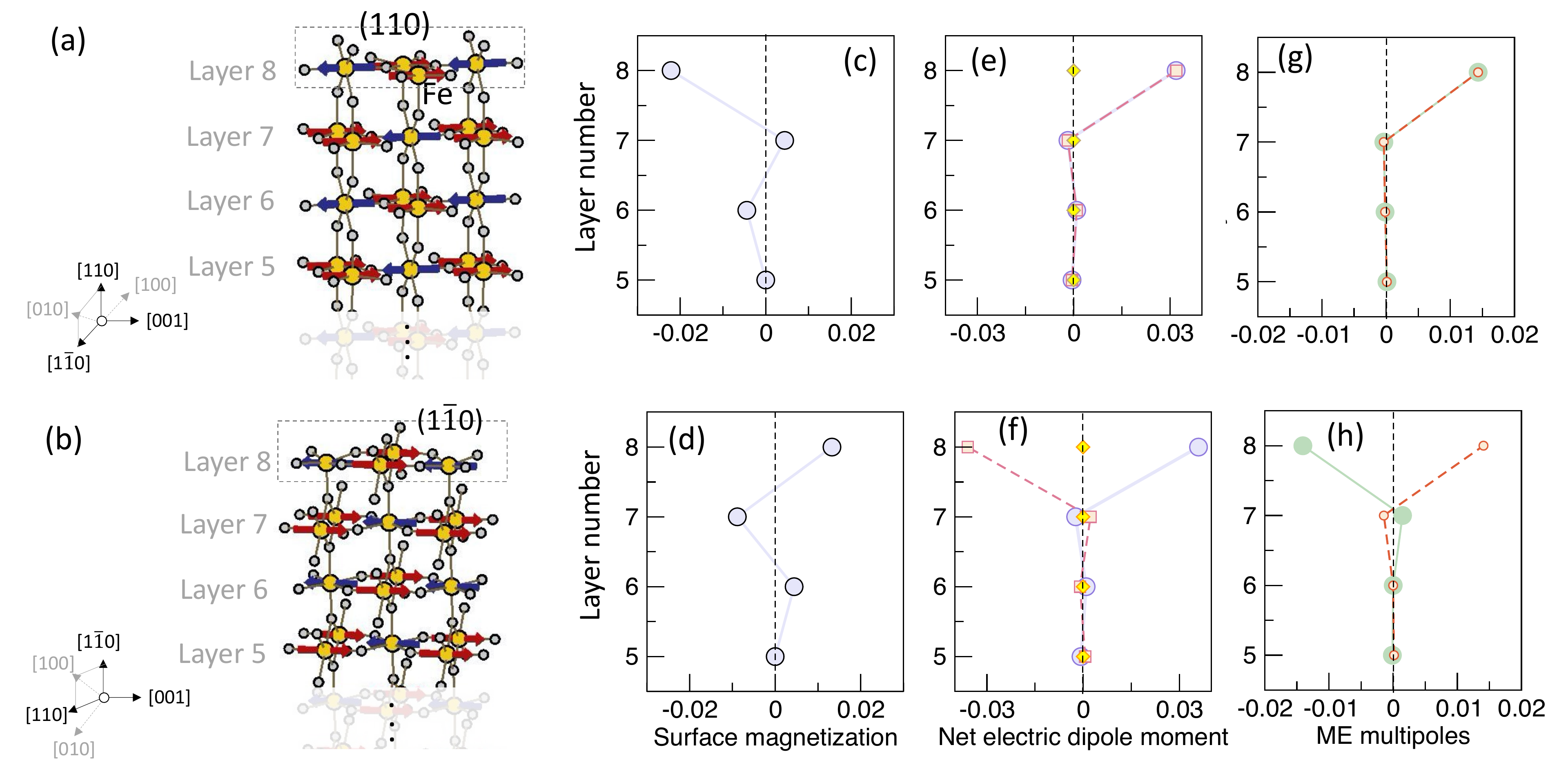}
\caption{
Emergence of multiferroicity for the (110) (top panel) and (1$\bar{1}$0) (bottom panel) surfaces. The (a) (110) and (b) (1$\bar{1}$0) surfaces of FeF$_2$. The arrows indicate the directions of the magnetic moments at the Fe ions. The dashed boxes highlight the individual surface layers. Since the slabs are inversion symmetric, only the upper surfaces are depicted. Computed layer-resolved surface magnetization (sum of the spin dipole moments along $\hat z$ at the Fe ions of a particular layer per unit surface area) in $\mu_B$/nm$^2$ for the (c) (110) and (d) (1$\bar{1}$0) surfaces. Layer-resolved net electric dipole moment (sum of the atomic site electric dipole moment at the individual ions of a particular layer) components $p_x$ (circles), $p_y$ (squares), and $p_z$ (diamonds) along the Cartesian directions in units of electronic charge for the (e) (110) and (f) (1$\bar{1}$0) surfaces. 
Computed layer-resolved ME multipole components (in units of $\mu_B$),
${\cal M}^{\rm surf}_{xz}$ (solid line) and ${\cal M}^{\rm surf}_{yz}$ (dashed line) 
for the (g) (110) and (h) (1$\bar{1}$0) surfaces. The emergence of ME multipoles, a net magnetic dipole, and an electric dipole moment is apparent at both surfaces.}
\label{fig3}
\end{figure*}

To compute the surface properties of FeF$_2$, we construct two separate slabs with $(110)$ and $(1\bar{1}0)$ surfaces, each containing eight Fe layers (see Fig. \ref{fig3}a and b). We constrain the in-plane lattice constants of the slabs to the relaxed bulk values $a=b=4.7803$ \AA~ 
and relax the internal coordinates and out-of-plane lattice constants, including approximately 15 \AA ~of vacuum along the surface normal. 
All calculations \cite{SM} are conducted without SOC. Both slabs maintain ${\cal I}$ symmetry, therefore we show the properties of only half of the layers in Fig. \ref{fig3}. We note that the central (fourth and fifth) layers of the slab have zero net ME multipole, net magnetic, and electric dipole moments, retaining the bulk-like behavior (see Fig. \ref{fig3}). 

First we note a net magnetic dipole moment along $[00\bar{1}]$ of the $(110)$ surface layers as shown in Fig. \ref{fig3}c, in agreement with previous theoretical predictions and calculations \cite{Munoz2015,Sophie2023}. 
Interestingly, we note that the sign of the net magnetic dipole moment switches to $[001]$ for the $(1\bar{1}0)$ surface [see Fig. \ref{fig3}d], consistent with the switching of the spin splitting in the bulk band structure. The two Fe ions become inequivalent at the surface, and, consequently, the opposite magnetic dipole moments of the Fe ions do not cancel each other, explaining the net magnetic dipole moment at the surface. 

It is interesting to point out that the emergence of surface magnetization can also be interpreted from the cross-section of the bulk anisotropic magnetization density, dictated by the magnetic octupoles. As shown in Fig. \ref{fig2}a, the magnetization density around the two Fe ions is inequivalent with the central Fe ion contributing more than the corner Fe ion at the (110) surface, leading to an uncompensated net magnetization, consistent with our finding as discussed above. Interestingly, an opposite situation occurs for the $(1\bar{1}0)$ surface cross-section of the bulk anisotropic magnetization density with the corner Fe ion contributing more than the central Fe ion. This results in a reversal of the net magnetization due to the opposite spin polarization of the Fe ions, corroborating our findings.

Next, we see that both FeF$_2$ surfaces exhibit a net electric dipole moment (summing the local electric dipole moments at the Fe and F sites) along the surface normal as depicted in Fig. \ref{fig3} (e) and (f). As seen from the figure, for the $(110)$ [$(1{\bar1}0)$] surface it is directed along $[110]$ ($[1{\bar1}0]$). 
The surfaces therefore have co-existing net magnetic moment and net electric dipole moment, and so can be classified as multiferroic. Interestingly, the magnetic and electric dipole moments are oriented perpendicular to each other, characteristic of a ME toroidal moment and an off-diagonal ME quadrupole.

Having established the coexistence of ferroically ordered magnetic and electric dipoles at the surfaces, we next analyze the next order ME multipoles. We find that 
both the $(110)$ and $(1\bar{1}0)$ surfaces exhibit a ferro-type ordering of ME multipole components, including ME quadrupole moments ${\cal Q}^{\rm ME}_{xz}$, ${\cal Q}^{\rm ME}_{yz}$, and ME toroidal moment components $t^{\rm ME}_{x}$, $t^{\rm ME}_{y}$, in contrast to the bulk where the ME multipoles vanish at the Fe site. Notably, for the $[110]$ surface, ${\cal Q}^{\rm ME}_{xz}$ and ${\cal Q}^{\rm ME}_{yz}$ have equal magnitudes and signs, while $t^{\rm ME}_{x}$ and $t^{\rm ME}_{y}$ have equal magnitudes but opposite signs. This leads to ${\cal M}^{\rm surf}_{xz} = ({\cal Q}^{\rm ME}_{xz}+t^{\rm ME}_{y})$ = ${\cal M}^{\rm surf}_{yz} = ({\cal Q}^{\rm ME}_{yz}-t^{\rm ME}_{x})$, as shown in Fig. \ref{fig3}(g). Conversely, the $[1\bar{1}0]$ surface shows ${\cal Q}^{\rm ME}_{xz}=-{\cal Q}^{\rm ME}_{yz}$ and $t^{\rm ME}_{x}=t^{\rm ME}_{y}$, leading to ${\cal M}^{\rm surf}_{xz} = -{\cal M}^{\rm surf}_{yz}$ (see Fig. \ref{fig3}(h)).

Since by symmetry the allowed non-zero elements of the linear ME response tensor correspond to the non-zero components of the ME multipole \cite{Spaldin2013}, our calculated ME multipole elements ${\cal M}_{xz}$ and ${\cal M}_{yz}$ should give rise to two non-zero response components $\alpha_{xz}$ and $\alpha_{yz}$.
Furthermore, since ${\cal Q}^{\rm ME}_{xz}={\cal Q}^{\rm ME}_{yz}$ and $t^{\rm ME}_{x}=-t^{\rm ME}_{y}$ for the $[110]$  surface layer, we expect $\alpha_{xz}=\alpha_{yz}$. On the other hand, for the $[1\bar{1}0]$ surface, we have 
$\alpha_{xz}=-\alpha_{yz}$. Thus, from the computation of the ME multipoles, we predict a change in magnetic moment along the $z$ direction as a linear response to an applied electric field along $x$ or $y$ direction and vice-versa in the $(110)$ and $(1\bar{1}0)$ surface layers.

\begin{figure}[t]
\includegraphics[width=\columnwidth]{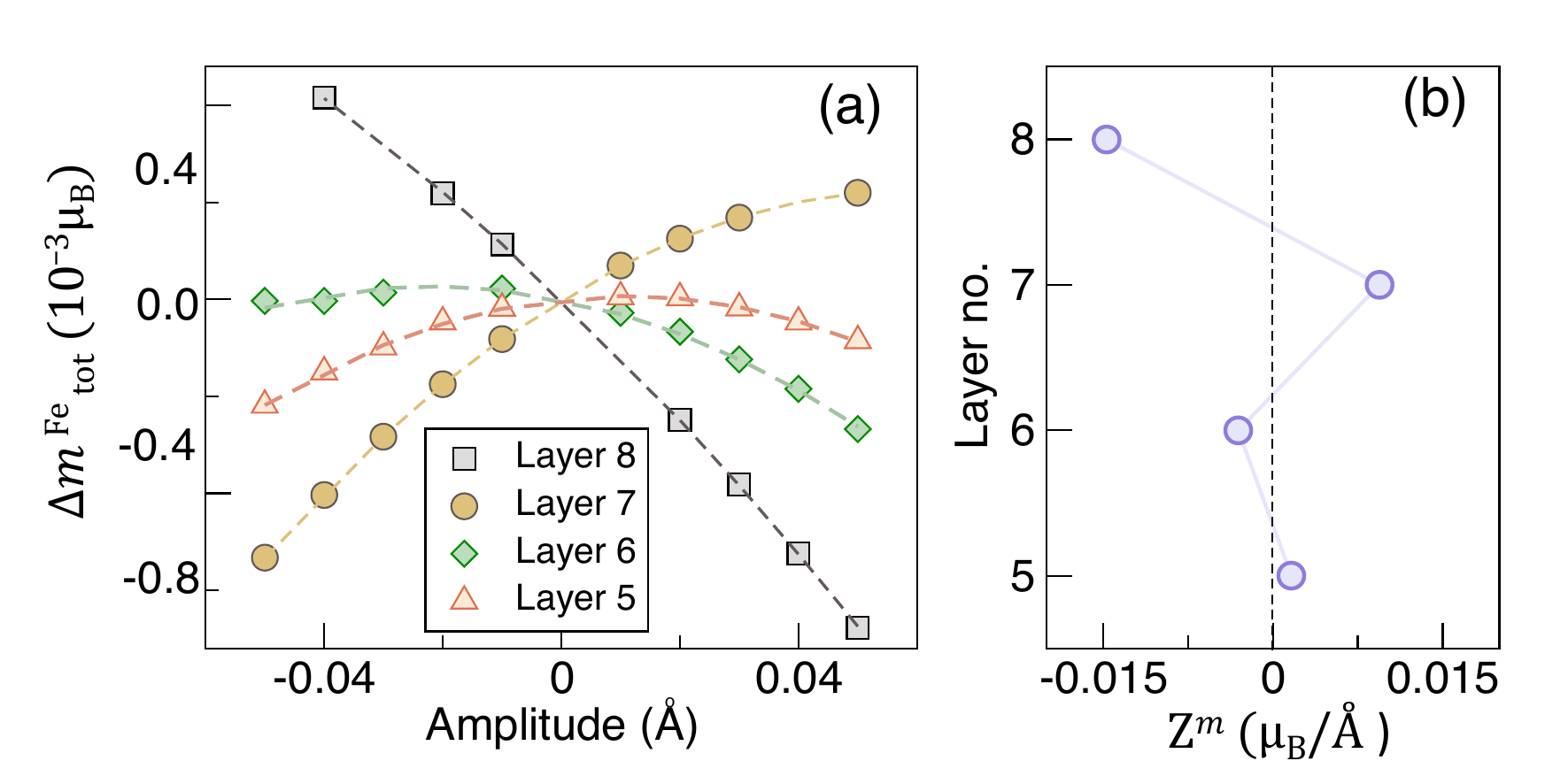}
\caption{(a) Computed total Fe magnetic moment ${\Delta} {m}^{z,\text{Fe}}_{\text{tot}}$ in the absence of SOC as a function of the displacement amplitude of Fe atoms in each layer. The dashed lines represent a second-order polynomial fit to the data points (shown as squares). (b) Layer-resolved DMC $Z^m$, extracted from the linear term of the polynomial fit in (a). 
}
\label{fig4}
\end{figure}

Next, we confirm the presence of the surface linear ME response by explicitly computing the corresponding lattice-mediated contribution, i.e., the net magnetic moment $m_j$ induced by the displacements $u_i$ of specific atoms $\alpha$, generated by an applied electric field. This is quantified by the \textit{dynamical magnetic charges} (DMCs)\cite{ye-prb2014}, defined as $Z^m_{\alpha, ij} = \frac{\partial m_{j}}{\partial u_{\alpha, i}}$. Here $i$ and $j$ denote the Cartesian directions. To compute the DMCs, we displace the Fe atoms in each layer along the [110] direction and compute the resulting change ${\Delta} {m}^{\text{Fe}}_{\text{tot}}$ in the total Fe spin moment for the entire slab (see Ref.~\cite{SM} for details). The computed ${\Delta} {m}^{\text{Fe}}_{\text{tot}}$ as a function of the displacement amplitude in each layer is shown in Fig. \ref{fig4}(a). Fitting of the computed data in Fig. \ref{fig4}(a) with a second-order polynomial (indicated in dashed lines) shows the presence of a combined linear and quadratic magnetic response to the displacement, analogous to an applied electric field in the direction perpendicular to the layer. We find that the linear response corresponding to the DMCs, which is our focus here, is most pronounced at the surface and then gradually decreases as we move towards the bulk-like inner layer, at which it becomes vanishingly small (see Fig. \ref{fig4}(b)). These findings are consistent with our computed layer ME multipoles and the magnetic point group $m'm2'$ symmetry of the surface, confirming the emergence of surface ME response even without SOC. While a surface ME effect has been noted previously in ferromagnets \cite{Chupis1994,Duan2008} or driven by SOC \cite{Eliseev2010}, this is to our knowledge the first prediction of the surface ME effect that occurs in an AFM without requiring SOC. 

We note that since the ME multipoles, and hence the ME response, are odd under inversion, they have opposite signs for the two surface layers of a given slab. Also, for the opposite magnetic domain with the flipped spin moments, our calculation shows that the surface magnetic dipoles and the ME multipoles switch sign, as expected due to their absence of $\cal T$ symmetry, while the direction of the surface electric dipoles remains the same.  

{\it Bulk-boundary correspondence--}
Having established the presence of the surface linear ME effect, we now show that it is also linked to the bulk magnetic octupoles of FeF$_2$. For this, we first note that as discussed above the  global and local bulk magnetic octupoles in FeF$_2$ give rise to a net magnetization along $\hat z$ as a second order response to an applied electric field of the form ${\vec E}\equiv(E_x, E_y,0)$, viz., 
\begin{eqnarray} \nonumber \label{octupole}
&& ~~\delta m_z \propto \sum_{j,k=x,y} {\cal O}_{zjk} E_j E_k \\ \nonumber
= &&(\bar{{\cal O}}_{zxx} E_x +{\cal O}_{zyx} E_y) E_x + (\bar{{\cal O}}_{zyy} E_y +{\cal O}_{zxy} E_x) E_y. \\
\end{eqnarray}
The $(110)$ surface has an intrinsic ${\vec E}$ field of the same form as above with $E_x=E_y$. Consequently, the intrinsic ${\vec E}$ field leads to a change in magnetization $\delta m_z$ that can be recast into surface ME multipole components ${\cal M}^{\rm surf}_{xz}$ and ${\cal M}^{\rm surf}_{yz}$, which correlate respectively to the linear ME response components $\alpha^{\rm surf}_{xz}$ and $\alpha^{\rm surf}_{yz}$ of the surface,
\begin{eqnarray} \label{MEmultipole}
\delta m_z \propto ({\cal M}^{\rm surf}_{xz} E_x + {\cal M}^{\rm surf}_{yz} E_y) .
\end{eqnarray}
Comparing Eqs. (\ref{octupole}) and (\ref{MEmultipole}), we obtain
\begin{eqnarray}  \nonumber  \label{}
{\cal M}^{\rm surf}_{xz} && \propto  (\bar{{\cal O}}_{zxx} E_x +{\cal O}_{zyx} E_y), \\ 
{\cal M}^{\rm surf}_{yz} && \propto  (\bar{{\cal O}}_{zyy} E_y +{\cal O}_{zxy} E_x).
\end{eqnarray}
Further, using  $\bar{{\cal O}}_{zxx}=\bar{{\cal O}}_{zyy}$ and ${\cal O}_{zxy}={\cal O}_{zyx}$ for FeF$_2$, and  $E_x=E_y$ for the $[110]$ surface, we obtain ${\cal M}^{\rm surf}_{xz}={\cal M}^{\rm surf}_{yz}$, consistent with our computed ME multipoles. In contrast, for the $[1\bar{1}0]$ surface $E_x=-E_y$ leads to ${\cal M}^{\rm surf}_{xz}=-{\cal M}^{\rm surf}_{yz}$ in agreement with our finding. Correspondingly, the surface ME response components are dictated by the non-zero bulk magnetic octupole components, and, hence, also occur without SOC. Since the surface ME response components, in turn, dictate which specific surfaces have a net magnetization, our work explains the surface magnetization for the (110) surface while its absence for the (001) surface \cite{Lapa2020, Munoz2015}.

It is important to point out that all $d$-wave spin split AFMs have a ferrotype ordering of magnetic octupoles, and when these octupoles occur in the absence of SOC, the bulk spin splitting is non-relativistic \cite{BhowalSpaldin2024}. As illustrated above, such bulk magnetic octupoles lead to ME multipoles at the surface and thus give rise to a linear ME effect which is nonrelativistic due to the nonrelativistic origin of the responsible multipoles. 
Thus, the non-relativistic surface linear ME effect is general to all $d$-wave NRSS materials. Furthermore, the
correspondence between the bulk octupole and surface
ME multipole leads to an interesting correlation between
the momentum directions $[110]$ and $[1\bar{1}0]$ of spin split-
ting in the bulk and the surface planes $(110)$ and $(1\bar{1}0)$,
exhibiting the ME effect.

Our work highlights the importance of bulk multipoles for
engineering interface properties and predicting emergent
phenomena under external perturbations.  Specifically, we
provide a framework for exploiting bulk magnetic octupoles
to achieve surface multiferroicity, with coexisting net
magnetization and polarization as well as a linear ME
response, from non-multiferroic bulk system.
We hope that our work motivates experimental studies to
confirm the predicted behavior, for example using nitrogen-vacancy magnetometry \cite{Appel2019}, surface magneto-optical Kerr measurements \cite{Qiu2000}, and magnetic force microscopy \cite{Wiesendanger2009} to measure the predicted
surface magnetization and its electric field-induced changes in FeF$_2$,
as well as to identify additional emergent surface behaviors
driven by ferroically ordered bulk multipoles.

\section*{Acknowledgements}
The authors thank David Vanderbilt for valuable discussions. 
This work was supported by the ERC under the EU’s Horizon 2020 Research and Innovation Programme grant No 810451 and by the ETH Zurich. SB acknowledges funding support from the Industrial Research and Consultancy Centre (IRCC) Seed Grant (RD/0523-IRCCSH0-018) and the INSPIRE research grant (project code RD/0124-DST0030-002). AU acknowledges support from the Abrahams Postdoctoral Fellowship of the Center for Materials Theory at Rutgers University. Computational resources were provided by ETH Zurich's Euler cluster, the Swiss National Supercomputing Centre, project ID eth3, and by the Beowulf cluster at the Department of Physics and Astronomy of Rutgers University.

\bibliography{LNPO}

\vspace{1 cm}

\begin{center}
\large\bf{Supplementary Material for Emergent surface multiferroicity} 
\end{center}

\section{Computational Details}

\begin{figure}[h]
\includegraphics[width=\columnwidth]{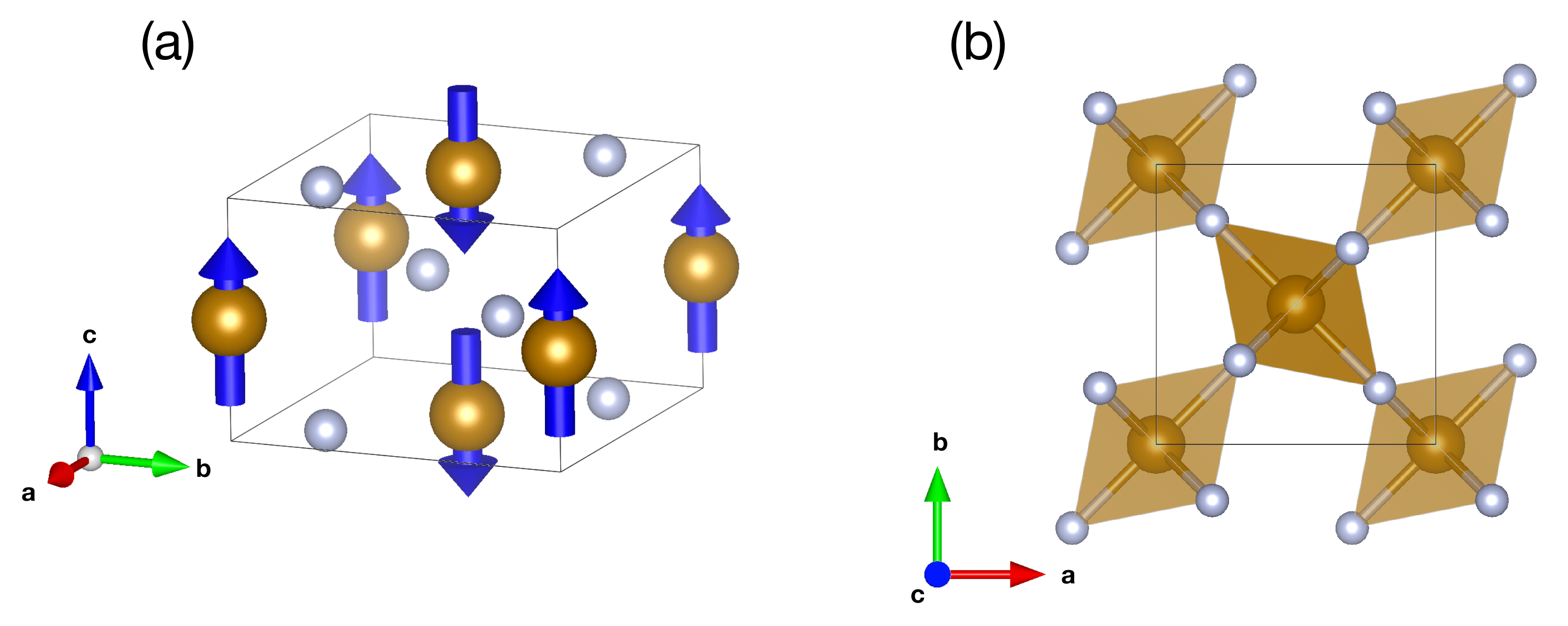}
\caption{Crystal structure of bulk FeF$_2$. Gold and grey balls identify Fe and F atoms, respectively. (a) 3D view, with magnetic moments indicated by blue arrows. (b) Top view, with FeF$_6$ octahedra highlighted.}
\label{figs1}
\end{figure}

All calculations presented in this manuscript were performed using density functional theory (DFT) within the plane-wave-based projector augmented wave (PAW) method \cite{Bloch1994, Kresse1999}, as implemented in the Vienna ab initio simulation package (VASP) \cite{Kresse1993, Kresse1996}. We employed the generalized gradient approximation with the Perdew-Burke-Ernzerhof (PBE) functional \cite{Perdew1996}, including a Hubbard $U$ correction of $6$ eV and a Hund's exchange $J$ of $0.95$ eV for the correlated Fe-$d$ states \cite{Sophie2023}. The spin-orbit interaction was not included in these calculations. 
The VASP PAW potentials for Fe-sv ([Ne]$3s^23p^63d^74s^1$) and F ([He]$2s^22p^5$) are used.

\begin{figure*}[t]
\includegraphics[width=0.9 \textwidth]{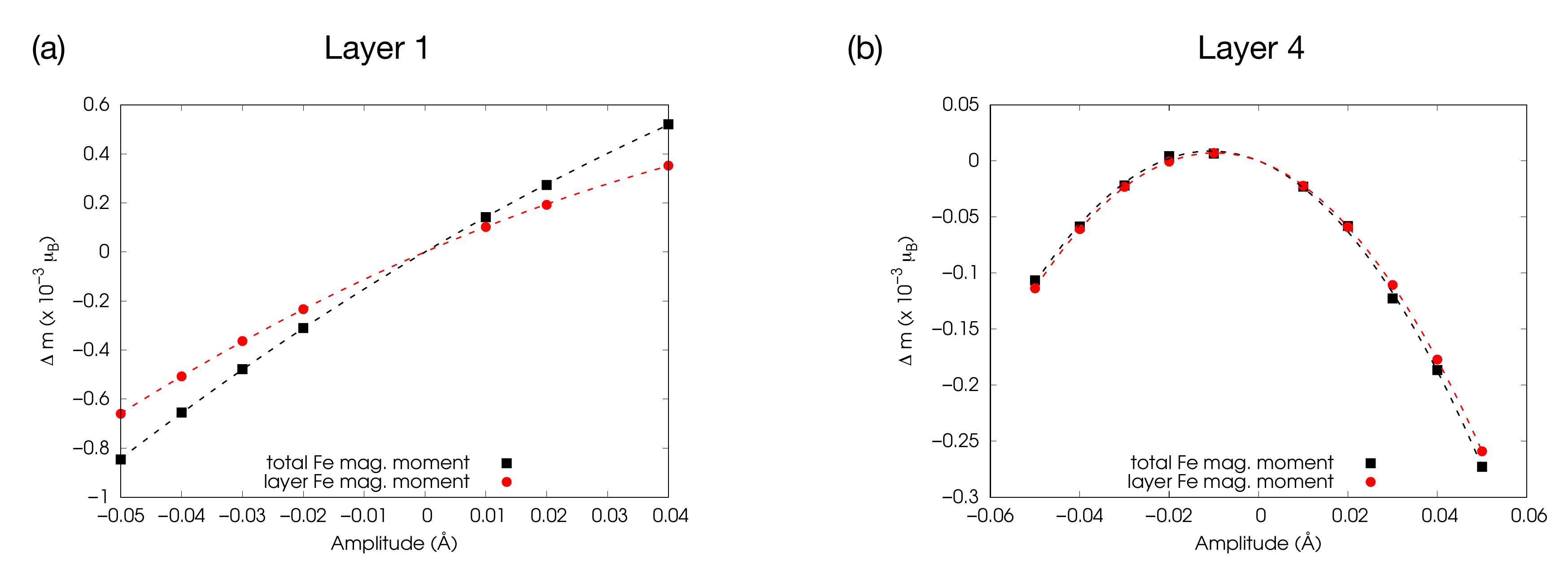}
\caption{Comparison between the sum of the magnetic response of the Fe ions in entire slab and the sum of the magnetic response of the Fe ions of the displaced layer. Results obtained by displacing (a) a surface layer (layer 1) and (b) a central bulk-like layer (layer 4). Dashed lines identify the fitting curve.}
\label{figs2}
\end{figure*}

For bulk FeF$_2$, the crystal structure of which is shown in Fig. \ref{figs1}, self-consistency was achieved with a 700 eV energy cut-off and a $6\times 6\times 9$ k-point grid. To explore surface properties, we constructed two slabs with eight Fe layers each, one with $(110)$ and another with $(1\bar{1}0)$ surfaces. The in-plane lattice constants of the slabs were constrained to the relaxed bulk values of $a=b=4.7803$ \AA. The internal coordinates were relaxed until the Hellmann-Feynman forces on each atom were less than 0.01 eV/\AA. Approximately 15 \AA\ of vacuum was introduced along the surface normal. Self-consistent calculations for the slab were carried out with an 800 eV energy cut-off and a $6\times 12\times 1$ k-point grid.

\begin{figure}[h]
\includegraphics[width=\columnwidth]{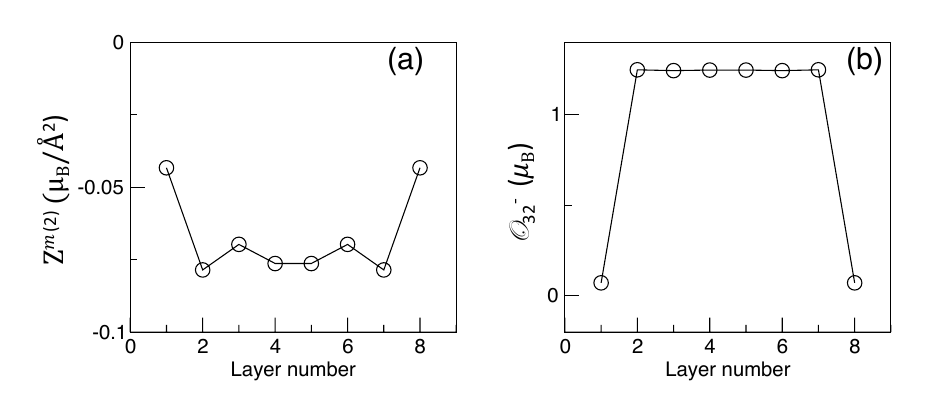}
\caption{(a) Layer-resolved second-order DMC $Z^{m (2)}$. (b) The variation of the magnetic octupole component ${\cal O}_{32}^{-}$ as a function of layer number.}
\label{figs3}
\end{figure}

The atomic-site multipoles at the Fe ions were calculated by decomposing the density matrix $\rho_{lm,l'm'}$, obtained from DFT, into irreducible (IR) spherical tensor components $w^{k p r}_t$ \cite{Cricchio2010, Spaldin2013}. Here, $k$, $p$, and $r$ represent the spatial index, spin index ($p = 0$ for charge, $p = 1$ for magnetic multipoles), and tensor rank ($r \in { | k - p |, | k - p | + 1, \dots, k + p }$), respectively, while $t \in { -r, -r + 1, \dots, r }$ labels the tensor components. The atomic-site magnetoelectric (ME) multipoles correspond to $k=1$, $p=1$, and $r=0, 1, 2$, representing the ME monopole, toroidal moment, and quadrupole moment, respectively. Since the ME tensor moments $w^{1 1 r}_t$ are odd under inversion, only terms  $l \ne l'$ (with $l + l'$ odd) in the density matrix contribute to these. Similarly, magnetic octupoles are defined by $k=2$, $p=1$, and $r=1, 2, 3$. The IR components of the magnetic octupole corresponding to $r=1, 2, 3$ denote respectively the moment of toroidal moment $t_i^{(\tau)}$, the toroidal quadrupole moment ${\cal Q}^{(\tau)}_{ij}$, and the totally symmetric components ${\cal O}_{lm}$ \cite{Urru2022}.  Unlike the ME multipoles, magnetic octupoles are inversion symmetric and only those terms in the density matrix $\rho_{lm,l'm'}$, for which $l+l'$ is even, contribute to the atomic-site magnetic octupoles.

Dynamical magnetic charges (DMCs) were computed from self-consistent calculations of distorted slab structures, obtained by displacing the Fe atoms of a given layer along the direction perpendicular to the slab layers. This was done for each layer separately and for different amplitudes of the total distortion, ranging from $-0.05$ \AA \, to $+0.05$ \AA \. The layer DMCs were then computed as the linear term of the fit of the total induced Fe spin moment for the entire slab as a function of the amplitude, as explained in the main text. The self-consistent calculations needed to evaluate the DMCs were performed using the same parameters listed earlier for the slab ground-state calculation, and magnetism was treated within the collinear spin-polarized framework.

We note that, because the undistorted slab has inversion symmetry, the DMCs (and, in turn, the linear ME response) are in principle expected to vanish. However, since we are interested in the properties of the surface, which breaks inversion symmetry, the DMCs are inherently non-zero. In order to access the DMCs, we slightly move one F atom away from its position in such a way to break inversion symmetry in the slab.

\section{Additional results for dynamical magnetic charge}

\subsection{Total vs. layer magnetoelectric response}

The lattice-mediated contribution to the ME response is the net magnetic moment induced by the ionic displacements generated by an applied electric field. To simplify the calculation, the total ME response can be broken down into the contributions due to the displacements of each single atom, which can be obtained from the DMCs \cite{ye-prb2014}. Since we are interested in the layer-resolved ME response, we compute the layer DMC by displacing the Fe atoms of a given layer, and computing the total Fe magnetic moment for the entire slab. It is important to point out that although there is a net magnetic dipole moment at the surface layers, the total magnetic moment (including all the atomic sites and the interstitial regions) of the slab remains zero within the collinear spin-polarized calculations in the absence of spin-orbit interaction. This is also supported from the sign change of the magnetic dipole moments at the layer adjacent to the surface. 

We expect the magnetic response to be dominated by the magnetic moment of the Fe atoms of the displaced layer. This is confirmed by Fig. \ref{figs2}, where we compare the total magnetic response of the Fe atoms for the entire slab with that of the Fe atoms of the displaced layer only. The two responses match almost exactly if we displace one of the central layers (Fig. \ref{figs2}(b)), while they are slightly different when displacing a surface layer, although still dominated by the Fe atoms of the surface layer.    


\subsection{Second-order dynamical magnetic charge}

As shown in the main text (Fig. 3), the magnetic response shows both a linear and a quadratic component in the amplitude of the displacement. For completeness, in Fig. \ref{figs3}(a) we show the second-order contribution, namely $Z^{m (2)}_{\alpha, ijk} = \frac{\partial^2 m_{k}}{\partial u_{\alpha, i} \partial u_{\alpha, j}}$, called the second-order DMC \cite{Urru2022}, for the different layers. We note that $Z^{m (2)}_{\alpha, ijk}$ and the computed layer-resolved magnetic octupole component ${\cal O}_{32}^-$, shown in Fig. \ref{figs3}(b), remain relatively constant in the inner layers and then decrease to lower values at the surface.

\end{document}